\documentclass[aps,prb,article,twocolumn,showpacs,longbibliography,preprintnumbers,amsmath,amssymb,superscriptaddress]{revtex4-2}

\usepackage{amssymb,amsmath,latexsym,mathrsfs}
\usepackage{graphicx}
\usepackage{physics}
\usepackage{mathtools}
\usepackage{bm}
\usepackage{bbold}
\usepackage[usenames,dvipsnames]{color}
\usepackage[breaklinks,colorlinks,urlcolor=magenta,citecolor=magenta,linkcolor=magenta]{hyperref}
\usepackage{xcolor}

\begin{document}

\title{Entropy from Entanglement in Quantum State Reduction}

\author{Lisa Lenstra}
\email{e.i.lenstra@uva.nl}
\affiliation{Institute for Theoretical Physics Amsterdam,
University of Amsterdam, Science Park 904, 1098 XH Amsterdam, The Netherlands}
\affiliation{QuSoft, Science Park 123, 1098 XG Amsterdam, The Netherlands}
\author{Jasper van Wezel}
\email{vanwezel@uva.nl}
\affiliation{Institute for Theoretical Physics Amsterdam,
University of Amsterdam, Science Park 904, 1098 XH Amsterdam, The Netherlands}


\begin{abstract}
The Von Neumann entropy of reduced states is a measure of bipartite entanglement. Despite its name, the entanglement entropy cannot by itself be used as a resource for creating thermodynamic heat flows. In order to extract heat from an entangled pure state, it first needs to be converted into a stochastically mixed state by a process of quantum state reduction. Here we show that even in a system with only two degrees of freedom, for which bipartite entanglement is the sole form of entanglement available, the entanglement entropy cannot be converted into thermodynamic entropy in a one-to-one fashion. Moreover, we show that the stochastic dynamics which is necessarily present in any realistic model of quantum state reduction, allows for multiple definitions of entropy. We indicate why quantum state reduction does not allow construction of a perpetuum mobile, despite some measures of entropy evolving non-monotonically during its dynamics. Finally, we relate the different measures of entropy to the information they contain about quantum entanglement and extractable heat, and show that models of quantum state reduction based on physical, correlated stochastic driving forces give rise to observable thermodynamic signatures of quantum state reduction that can be unambiguously distinguished from environment-induced dephasing. 
\end{abstract}

\maketitle

\section{Introduction}
Quantum entanglement and thermodynamic entropy are closely related concepts. In fact, one of the most used measures for bipartite entanglement in quantum systems is known as the (reduced) Von Neumann entropy \cite{popescu,vedral,vidal,horodecki2}. However, to create a statistical ensemble with non-zero thermodynamic entropy starting from a collection of identical pure entangled quantum states, a physical operation is required. The quantum states need to be measured, in order to project them onto a set of unentangled basis states with Born rule probabilities. Only after such measurement does the ensemble contain thermodynamic entropy that can be used to, for example, create a flow of heat --using the second law-- or to do work --using the first law at constant internal energy \cite{giles,heredecki}. The reduction of a pure initial state to a true statistical mixture, rather than tracing a pure state over environmental degrees of freedom to obtain a mixed effective description, can be described by models of objective quantum state reduction \cite{CSL,Mertens2,jvw,Bassi}.

Several such models have been proposed as solutions to the measurement problem of quantum mechanics, but to date none have been experimentally verified or falsified, because of the extremely challenging conditions required for probing their predicted dynamics \cite{Bassi,bassi3,pearle3,Marshall,oosterkamp,vinante2017,Adler_exp}. Here, we show that the conversion of quantum entanglement into thermodynamic entropy follows different evolutions for distinct models of quantum state reduction. The generation of thermodynamic entropy thus contains signatures of the way that quantum measurement unfolds, and can be used to distinguish between different types of models for objective quantum state reduction. Moreover, because changes in thermodynamic entropy can be used to generate flows of heat or to do work, the creation of entropy during measurement can be used to constrain models of objective quantum state reduction by requiring that they do not allow any form of perpetuum mobile. We show that this requirement is satisfied for several large classes of models, despite some of them not conserving the average weight of wavefunction components (i.e. their dynamics not being a Martingale process \cite{mukherjee2,mukherjee3}), and some measures for entropy evolving non-monotonically during quantum state reduction.

Finally, we show that multiple, inequivalent ways of defining statistical ensembles are possible for entangled quantum systems undergoing measurement dynamics. We use these to show that regardless of the way in which quantum measurement works, the entanglement entropy in the initial state is never converted into thermodynamic entropy in a one-to-one fashion, indicating that entanglement and entropy are not interchangeable physical properties.

\section{Quantum state reduction}
All models for objective quantum state reduction propose generalisations of Schr\"odinger's equation, whose effect is unobservable at microscopic scales while it dominates the dynamics of macroscopic objects~\cite{Mertens2,jvw,mukherjee2,Bassi,bassi3}. The result on everyday scales is a fast dynamical reduction of generic quantum superpositions to classical states. Importantly, the proposed dynamics affects individual systems, without the need to average over any environmental degrees of freedom, which makes it distinct from environment-induced decoherence~\cite{jvw,Bassi,oosterkamp,adler}. To serve as a solution to the quantum measurement problem, the probability of obtaining any particular classical state in models of quantum state reduction needs to agree with Born's rule. This is only possible if the proposed dynamics in is both stochastic and non-linear \cite{Mertens2}.

The time evolution of states in models of objective quantum state reduction can be written in the generic form~\cite{mukherjee2}:
\begin{align}
    d\ket{\psi} &=\left( -i\hat{H}dt + d\hat{\mathcal{G}}_t(\psi,\xi^j)\right)\ket{\psi}. 
    \label{eq:general}
\end{align}
Here we used units in which $\hbar=1$, while $\hat{H}$ is the usual quantum mechanical Hamiltonian, and $d\hat{\mathcal{G}}_t$ denotes the introduced deviation from Schr\"odinger dynamics. In general, it is non-linear and stochastic, so that it depends on the instantaneous state $\ket{\psi}$ it acts on, as well as a collection of stochastically evolving variables $\xi^j(t)$. The precise form of $d\hat{\mathcal{G}}_t$ differs from one model for quantum state reduction to another~\cite{mukherjee2,bassi3}. To ensure that it has negligible effect in the microscopic limit, it is often assumed to be proportional to an extensive observable, such as mass or an order parameter \cite{jvw, Bassi, bassi3}.

In the specific case of quantum measurement with $\ket{\psi}$ representing a macroscopic system superposed over two classical states in its initial configuration, the leading order corrections to Schr\"odinger dynamics can be written in the form~\cite{Mertens2,Scipost,mukherjee2}:
\begin{align}
    d\ket{\psi} &=\left( -i\hat{H} + \left(J\langle\hat{\sigma}\rangle + G \xi(t) \right) \left(\hat{\sigma}-\langle\hat{\sigma}\rangle\right)  \right)\ket{\psi}dt. 
    \label{eq:two-state}
\end{align}
Here, we assumed that the evolution of the stochastic parameter $\xi(t)$ is itself independent of the quantum state, and $\hat{\sigma}$ is a Pauli-$\sigma_z$ operator acting on the basis of superposed classical states. It represents an extensive physical observable, so that its prefactor $J$ scales with system size. The expectation value $\langle\hat\sigma\rangle$ in the factor $\left(\hat{\sigma}-\langle\hat{\sigma}\rangle\right)$ ensures that the dynamics of $\ket{\psi}$ is norm-preserving, but does not affect the relative weights or phases of its components~\cite{mukherjee2}. The form of Eq.~\eqref{eq:two-state} is unique in the sense that other forms leading to Born rule statistics at late times necessarily contain higher-order nonlinearities \cite{Mertens2,Scipost}. 

The stochastic parameter $\xi(t)$ is often referred to as noise. It is temporally correlated, and obeys the independent stochastic dynamics $d\xi=-\xi dt/\tau + g(\xi)dW_t$~\cite{mukherjee2}. Here, $dW_t$ indicates the increment for a Wiener process and is sampled from a Gaussian probability distribution with standard deviation $\sqrt{dt}$. The parameter $\tau$ denotes the correlation time for the noise, while different choices for the function $g(\xi)$ define distinct models for the evolution of the noise. In the so-called white noise limit, where $\tau$ is going to zero, Eq.~\eqref{eq:two-state} becomes weakly equivalent (i.e. agreeing on probabilities) to a model for Continuous Spontaneous Localization (CSL)~\cite{CSL,mukherjee4}. However, given that all physical processes necessarily have non-zero correlation time~\cite{hida,papoulis, mukherjee4}, we here consider white noise only as the limit of a correlated process. In fact, we start by considering the opposite limit, of infinite correlation time. Given that $\xi(t)$ represents physical fluctuations, we can always consider the measurement dynamics of an object that is sufficiently large for it to reach a final state within a time shorter than the correlation time of the fluctuations~\cite{Mertens_core}. In that case, $\xi(t)$ is approximately constant during each implementation of the dynamics of Eq.~\eqref{eq:two-state}, but obtains a new, randomly sampled value for each consecutive implementation.

\begin{figure}[tb]
\begin{center}
\includegraphics[width=\columnwidth]{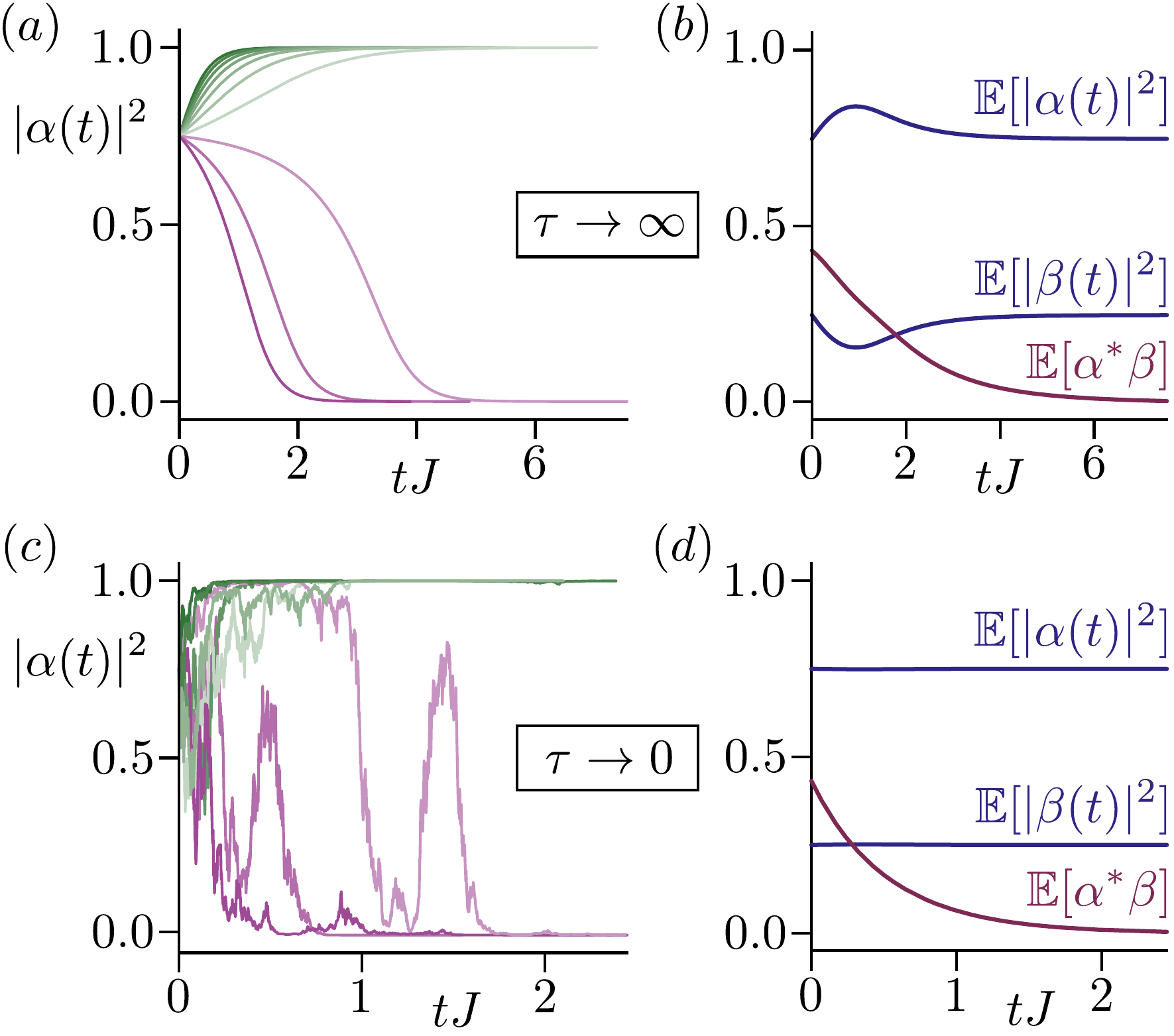}
\caption{\label{fig1} {\bf State evolution during quantum state reduction.} (a) The evolution of the state $\ket{\psi(t)}=\alpha(t) \ket{0} + \beta(t) \ket{1}$ under Eq.~\eqref{eq:two-state} with $J=G$ and $\hat{H}=0$ is shown for different values of the stochastic parameter $\xi$, sampled from a uniform distribution on $[-1,1]$. (b) The average weight of the wave function components and their overlap, indicating that the proportion of states ending in $\ket{0}$ at late times is equal to the value of $|\alpha|^2$ at $t=0$, in accordance with Born's rule~\cite{Mertens2,Scipost,mukherjee3}. (c) and (d) show the same as panels (a) and (b), but in the white noise limit of vanishing correlation time $\tau$ for the stochastic parameter, in which the average dynamics coincides with that of a CSL (continuous spontaneous localization) model~\cite{mukherjee4}.}
\end{center}
\end{figure}

To obtain Born rule statistics for the late-time states reached under the evolution defined by Eq.~\eqref{eq:two-state}, the amplitude $G$ of the stochastic term needs to be related to the strength $J$ of the non-linear term \cite{Mertens2,mukherjee2}. This can be seen as a type of fluctuation-dissipation relation, linking the tendency to drift towards a classical steady state to the strength of instantaneous stochastic fluctuations. In the $\tau \to \infty$ limit, the relation $J=G$ has been shown to result in the correct late-time statistics \cite{Scipost}. Fig.~\ref{fig1} shows some typical evolutions of the state $\ket{\psi(t)}=\alpha(t) \ket{0} + \beta(t) \ket{1}$ under the dynamics of Eq.~\eqref{eq:two-state} in both the infinite and zero correlation time limits, as well as their behavior averaged over different realizations of the stochastic parameter.

\section{Entropy and Entanglement}
To visualize the relation between entanglement and entropy, we consider an ensemble of $N$ initially identical entangled states $\ket{\psi} = \alpha \ket{00} + \beta \ket{11}$. Each state consists of two space-like separated quantum systems, accessible to agents called Alice and Bob respectively, such that $\ket{00} = \ket{0}_{\text{A}}\otimes\ket{0}_{\text{B}}$ and similarly for $\ket{11}$. We then consider the situation in which Alice performs a quantum measurement with a large but finite-sized measurement apparatus, while Bob does nothing. We assume a so-called strong measurement scenario as originally introduced by Von Neumann~\cite{vNeumann_strong}, in which a strong coupling between the measurement apparatus and Alice's quantum system causes them to become fully entangled in an arbitrarily short time, after which the measurement dynamics of Eq.~\eqref{eq:two-state} commences. In the initial ensemble, the state $\ket{0}_{\text{A}}$ can then be taken to denote a product of states for the measurement apparatus and the microscopic system being measured by Alice, so that $\ket{0}_{\text{A}}=\ket{0}_{\text{Apparatus A}}\otimes\ket{0}_{\text{System A}}$.

\begin{figure*}[tb]
\begin{center}
\includegraphics[width=\textwidth]{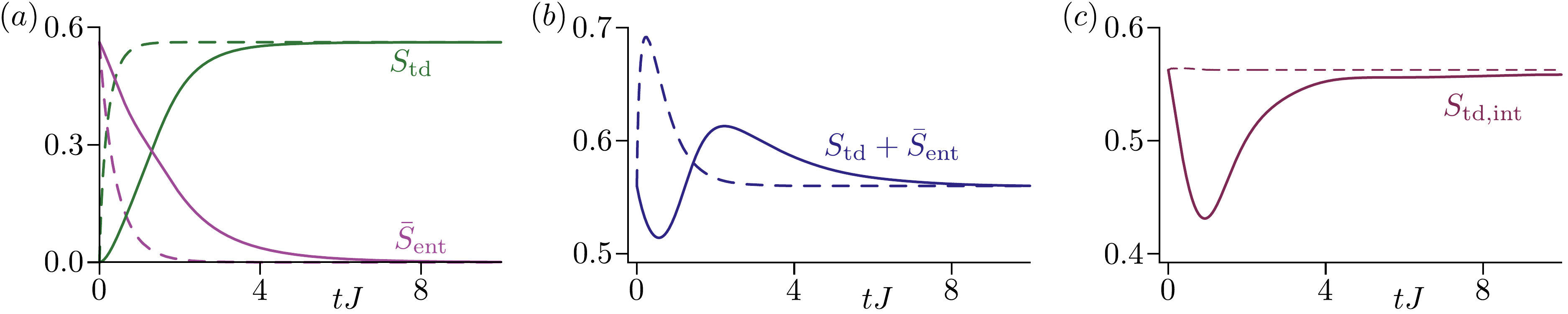}
\caption{\label{fig2} {\bf Evolution of entropy and entanglement during quantum state reduction.} (a) The evolution of the instantaneous thermodynamic entropy $S_\text{td}(t)$ and the average bipartite entanglement $\bar{S}_\text{ent}(t)$ during quantum state reduction. (b) Sum of the thermodynamic entropy and average bipartite entanglement during quantum state reduction. The sum is not constant, indicating that entanglement is not converted into thermodynamic entropy in a one-to-one fashion. (c) The evolution of the locally obtainable entropy $S_{\text{td,int}}(t)$ during quantum state reduction. Although the locally obtainable entropy is not a monotonic function of time, this cannot be used to obtain a decreasing value of the actually obtained thermodynamic entropy in any situation. In all panels, the lines show ensemble averages over $N=10^5$ implementations of the stochastic parameter, with the state initialized at $\alpha(0) = \sqrt{3/4}$. Solid lines indicate evolution according to the dynamics given by Eq.~\eqref{eq:two-state} with constant $\xi$, while dashed lines indicate the dynamics obtained from the white-noise limit of the same equation.
}    
\end{center}
\end{figure*}

Due to the measurement apparatus not being infinitely large, the measurement dynamics will take a nonzero time to reach a steady state representing a particular measurement outcome. We use Eq.~\eqref{eq:two-state} in the infinite correlation time limit with $J=G$ and $\hat{H}=0$ to describe the evolution of individual quantum states. For each state in the ensemble, a value of $\xi$ is randomly drawn from a uniform distribution on $[-1,1]$, and causes it to evolve from $\ket{\psi(0)}=\alpha(0)\ket{00}+\beta(0)\ket{11}$ to either $\ket{00}$ or $\ket{11}$ as a function of time. The ensemble as a whole can be described by the density matrix:
\begin{align}
\rho(t) &= \frac{1}{N} \sum_{j} \ket{\psi(\xi_j,t)}\bra{\psi(\xi_j,t)} \notag \\
&= \begin{pmatrix}
    \mathbb{E}_\xi \left[|\alpha(\xi,t)|^2\right] & \mathbb{E}_\xi \left[\alpha^*(\xi,t)\beta(\xi,t)\right]\\
    \mathbb{E}_\xi \left[\beta^*(\xi,t)\alpha(\xi,t)\right] & \mathbb{E}_\xi \left[|\beta(\xi,t)|^2\right]
\end{pmatrix}.
\label{eq:densitymatrix}
\end{align}
Here $\mathbb{E}_\xi$ indicates the ensemble average over $N$ implementations of the stochastic variable.

The instantaneous thermodynamic entropy contained in the ensemble described by $\rho$, which in conjunction with the second law can be converted into an observable flow of heat, is given by~\cite{giles,heredecki}:
\begin{equation}
     S_{\text{td}}(t) = - \text{Tr}(\rho(t) \ln \rho(t)).
     \label{eq:std}
\end{equation}
Here we use units in which the Boltzmann constant $k_{\text{B}}$ equals one. At the initial time $t=0$, the density matrix $\rho$ is a pure state, so that the thermodynamic entropy and extractable heat flow are zero. At late times, each quantum state in the ensemble has been reduced to either $\ket{00}$ or $\ket{11}$, with the distribution of final states obeying Born rule statistics~\cite{Mertens2,Scipost,mukherjee2}. The density matrix is then mixed and equal to $\rho=\text{diag}[|\alpha(0)|^2, |\beta(0)|^2]$, while the entropy tends to its maximal value of $S_{\text{td}}(t\to\infty) = - |\alpha(0)|^2 \ln |\alpha(0)|^2 - |\beta(0)|^2 \ln |\beta(0)|^2$ (see Fig.~\ref{fig2}a).

At the same time, we can consider the average value of the entanglement contained in the individual states making up the ensemble. For a single quantum state, the bipartite entanglement entropy (i.e. the reduced Von Neumann entropy) is given by~\cite{popescu,vedral,vidal,horodecki2}:
\begin{align}
S_{\text{ent}}(\xi,t) &= - \text{Tr}_{\text{A}}\left(\rho_{\text{A}}(\xi,t) \ln \rho_{\text{A}}(\xi,t)\right).
    \label{eq:pure_ent}
\end{align}
Here $\rho_\text{A} = \text{Tr}_{\text{B}}(\rho)$ is the reduced density matrix relevant for predicting expectation values of local observables on Alice's side, which is obtained by tracing over the states localized at Bob's position. The average bipartite entanglement entropy per state can then be defined for the entire ensemble as $\bar{S}_{\text{ent}}(t) = \mathbb{E}_{\xi}\left[ S_{\text{ent}}(\xi,t) \right]$.
Notice that $\bar{S}_{\text{ent}}$ constructed this way gives a measure of entanglement only, while first constructing the density matrix $\bar{\rho} = \mathbb{E}_\xi[ \rho ]$ for the statistical ensemble and then calculating $\text{Tr}_{\text{A}}\left(\bar{\rho}_{\text{A}} \ln \bar{\rho}_{\text{A}}\right)$ would result in a quantity containing contributions from both entanglement and classical (mixing) entropy. 

Starting from $N$ pure, entangled states, $\bar{S}_{\text{ent}}$ initially has the maximal value  $\bar{S}_{\text{ent}}(0)= - |\alpha(0)|^2 \ln |\alpha(0)|^2 - |\beta(0)|^2 \ln |\beta(0)|^2$, which equals $S_{\text{td}}(t\to\infty)$, the late-time value value for the thermodynamic entropy. At the end of the quantum measurement dynamics the ensemble contains unentangled states only and $\bar{S}_{\text{ent}}(t\to\infty)$ tends to zero, as shown in Fig.~\ref{fig2}a. Despite the commonly used name ``entanglement entropy'', and despite the initial value of $\bar{S}_\text{ent}$ being equal to the final value of $S_\text{td}$, the measurement dynamics does not convert entanglement into entropy in a one-to-one fashion. This is especially noticeable in the sum of thermodynamic entropy and average bipartite entanglement, shown in Fig.~\ref{fig2}b, which is not constant in time. In fact, this conclusion is not specific for the infinite correlation time limit considered here, and occurs even in the opposite, white-noise limit, in which Eq.~\eqref{eq:two-state} reduces to a CSL model~\cite{mukherjee4}, as indicated by the dashed lines in Figs.~\ref{fig2}ab.

More generally, for any model of objective state reduction, the reduced density matrix in Eq.~\eqref{eq:pure_ent} will equal $\rho_\text{A}(\xi,t)=|\alpha(\xi,t)|^2 \ket{0} \mskip -4mu \vphantom{\ket{1}}_{\text{A}}   \vphantom{\ket{1}}_{\text{A}} \mskip -5mu \bra{0} + |\beta(\xi,t)|^2 \ket{1} \mskip -4mu \vphantom{\ket{1}}_{\text{A}}   \vphantom{\ket{1}}_{\text{A}} \mskip -5mu \bra{1}$. The average bipartite entanglement can therefore always be written as:
\begin{align}
\bar{S}_{\text{ent}}(t) = &- \mathbb{E}_{\xi} \mskip -4mu \left[ \, |\alpha(\xi,t)|^2 \ln |\alpha(\xi,t)|^2 \right] \notag \\ &- \mathbb{E}_{\xi} \mskip -4mu \left[ \, |\beta(\xi,t)|^2 \ln |\beta(\xi,t)|^2 \, \right].
\label{eq:entgen}
\end{align}
The thermodynamic entropy meanwhile, can be written as $S_\text{td}(t) =  -x_+(t)\ln x_+(t) - x_-(t)\ln x_-(t)$, with $x_\pm$ the eigenvalues of the density matrix in Eq.~\eqref{eq:densitymatrix}:
\begin{align}
     x_\pm(t) =  -\frac{1}{2}\pm\frac{1}{2} &\left\{ \left(\mathbb{E}_\xi \mskip -4mu \left[ \, |\alpha(\xi,t)|^2-|\beta(t)|^2 \, \right] \right)^2 \right. \notag \\ 
     &\phantom{..} + \left. 4 \left| \, \mathbb{E}_\xi \mskip -4mu \left[ \, \alpha(\xi,t)\beta^*(\xi,t) \, \vphantom{.^2} \right] \, \right|^2 \right\}^{1/2}.
     \label{eq:EVstd}
\end{align}
Substituting the eigenvalues of Eq.~\eqref{eq:EVstd} into the entropy $S_\text{td}(t)$ yields an expression that is unrelated to Eq.~\eqref{eq:entgen} for the average bipartite entanglement. Except possibly at special, isolated instances of time, the sum of entanglement and entropy will therefore deviate from its initial value in any model for objective quantum state reduction.

The difference between entanglement and thermodynamic entropy observed here is consistent with recent observations that no ``second law of entanglement'' can exist~\cite{lami}. This is made explicit in the present case by the evolution of the sum shown in Fig.~\ref{fig2}b. Notice that this does not lead to any problems associated with violations of the second law of thermodynamics, however, because the thermodynamic entropy $S_{\text{td}}$ is a monotonously increasing function of time. Only thermodynamic entropy can be used to do work, while quantum entanglement cannot, which guarantees that during the evolution shown in Fig.~\ref{fig2} no work can be generated without external energy input.

\section{Interruptive measurement}
The measurement scenario can be extended by allowing the dynamics of quantum state reduction to be interrupted after a given time $t$ by an instantaneous projective measurement. We thus again consider Alice and Bob at time $t=0$ sharing $N$ entangled states of the form $\ket{\psi} = \alpha \ket{00} + \beta \ket{11}$, and Alice performing a quantum measurement with a large but finite-sized measurement apparatus, while Bob initially does nothing. We again consider the dynamics of individual quantum states in this interval to be given by Eq.~\eqref{eq:two-state}. Then, at time $t$, Bob instantaneously measures the state using an infinitely large apparatus. As a result, each state is projected onto the basis $\{ \ket{00}, \ket{11} \}$ at time $t$, with probabilities corresponding to their instantaneous values of $|\alpha(\xi,t)|^2$ and $|\beta(\xi,t)|^2$. The full density matrix is then given by:
\begin{align}
    \rho_{\text{int}}(t) = \begin{pmatrix}
    \mathbb{E}_{\xi} \mskip -4mu\left[\, |\alpha(\xi,t)|^2 \, \right] & 0\\
    0 & \mathbb{E}_{\xi} \mskip -4mu\left[ \, |\beta(\xi,t)|^2 \, \right]
\end{pmatrix}.
\label{eq:intdensitymatrix}
\end{align}    

The thermodynamic entropy of the ensemble after the instantaneous projection at time $t$ is defined by Eq.~\eqref{eq:std}, and becomes:
\begin{align}
    S_{\text{td,int}}(t) = &- \mathbb{E}_{\xi} \mskip -4mu \left[ \, |\alpha(\xi,t)|^2 \right] \ln \mathbb{E}_{\xi} \mskip -4mu \left[ \, |\alpha(\xi,t)|^2 \right] \notag \\ 
    &- \mathbb{E}_{\xi} \mskip -4mu \left[ \, |\beta(\xi,t)|^2 \right] \ln \mathbb{E}_{\xi} \mskip -4mu \left[ \,|\beta(\xi,t)|^2 \right].
    \label{eq:SN}
\end{align}
The quantity $S_{\text{td,int}}$ indicates the thermodynamic entropy that could be obtained locally by Bob upon performing an idealized, instantaneous measurement that converts all entanglement remaining at time $t$ into entropy. We therefore refer to it as the locally obtainable entropy. It can be used by Bob to locally do work or to cause local heat flows.

As indicated in Fig.~\ref{fig2}c, the evolution of the locally obtainable entropy is not monotonic if Alice uses a stochastic parameter with non-zero correlation time, while it becomes constant in the white-noise limit. The locally obtainable entropy $S_{\text{td,int}}$ contains contributions from both the instantaneous thermodynamic entropy $S_{\text{td}}$ and the conversion of any remaining entanglement $\bar{S}_{\text{ent}}$ into entropy. Notice however that it is not simply the sum of these (shown in Fig.~\ref{fig2}b). This is because Bob's instantaneous measurements always yield one of two possible outcomes, whilst before measurement there are many distinctly entangled states, resulting from many different implementations of the stochastic variable $\xi$. Bob's measurements thus convert entangled states into a classical mixture, but they also reduce the total number of distinct states in the ensemble.

The fact that the locally obtainable entropy may evolve non-monotonically and initially decreases as a function of time, does not indicate a violation of the second law of thermodynamics. What decreases is the potentially obtainable entropy, rather than any actually obtained quantity. That is, even though an instantaneous projective measurement at $t\approx 1/J$ results in a lower entropy than a measurement at $t \approx 0$ would have yielded, the measurement at $t\approx 1/J$ can only be performed if the one at $t \approx 0$ did not take place. The obtained entropy being lower than a value that was never realized can not be used to power a perpetuum mobile or circumvent the laws of thermodynamics.

Notice that these results still hold if Bob decides to projectively measure only a fraction of the $N$ states at his disposal. In that case, we can define two separate ensembles: one being reduced to the density matrix of Eq.~\eqref{eq:intdensitymatrix} at the time of Bob's measurement, while the other keeps evolving according to Eq.~\eqref{eq:densitymatrix}. In both ensembles, the thermodynamic entropy associated with their actual density matrix increases monotonically as a function of time. Finally, the entropy that might be gained from statistically mixing the two ensembles is strictly non-negative, and can therefore not be used to achieve any decrease in the total thermodynamic entropy~\cite{prigogine}.

\section{Dephasing}
That the locally obtainable entropy evolves non-monotonically under the dynamics of Eq.~\eqref{eq:two-state} driven by physical, correlated noise, is a direct consequence of the diagonal elements in the density matrix of Eq.~\eqref{eq:intdensitymatrix} not being constant in those models. Similarly, the locally obtainable entropy is constant in models such as CSL, because these are constructed with the explicit requirement that $\mathbb{E}_\xi [|\alpha(\xi,t)|^2]$ is constant, i.e. that their dynamics is a Martingale process \cite{bassi3,pearle,CSL}. As noted above, the averaged dynamics of Eq.~\eqref{eq:two-state} coincides with a CSL model in the white noise limit~\cite{mukherjee4}. Any observation of a non-monotonically evolving locally obtainable entropy may therefore be interpreted as an experimental fingerprint of objective quantum state reduction driven by stochastic noise with a non-zero correlation time. For a class of models for objective quantum state reduction, the locally obtainable entropy thus provides a thermodynamic fingerprint that does not rely on the observation of interference patterns and hence opens an alternative experimental window on the dynamics of quantum state reduction.

Any experimental signature of objective quantum state reduction, however, is only useful if it can be distinguished from the effects of environment-induced dephasing~\cite{jvw,Bassi,oosterkamp,adler}. This is not possible for the locally obtainable entropy in CSL models, including Eq.~\eqref{eq:two-state} in the limit of vanishing correlation time, because these constitute an unraveling of the master equation for pure dephasing~\cite{Bassi, CSL}. That is, these models are constructed precisely such that their predicted averages for any system are indistinguishable from the local expectation values that would result from the same system undergoing pure Schr\"odinger evolution while coupled to an unobserved environment.

To be explicit, consider again the interruptive measurement dynamics of the previous section, but with Alice allowing her qubit to entangle with an environment rather than imposing any objective quantum state reduction. In that case, the state at time $t$ is given by:
\begin{align}
\ket{\psi_\text{env}(t)} &= e^{-\frac{i}{\hbar}\hat{H}t} \left( \alpha \ket{00} + \beta \ket{11} \right) \ket{\text{env}} \notag \\ 
&= \alpha \ket{00\;\text{env}_0(t)} + \beta \ket{11\;\text{env}_1(t)}.
\label{eq:stateenv}
\end{align}
Here, $\hat{H}$ is the total Hamiltonian of both qubits, the environment, and their interactions, while $\ket{\text{env}_0(t)}$ signifies the state that the environment would reach at time $t$ if it started from $\ket{00}\otimes\ket{\text{env}}$. We consider a pure dephasing process, in which the environment does not affect the state of the qubits. Notice that Eq.~\eqref{eq:stateenv} differs from the state $\ket{\psi(\xi_j,t)} = \alpha(\xi_j,t) \ket{00} + \beta(\xi_j,t)\ket{11}$, which we obtained using the dynamics of Eq.~\eqref{eq:two-state}. Nevertheless, in the limit of vanishing correlation time, both can be used to obtain the same reduced density matrix:
\begin{align}
\Tr_{\text{env}} \left[ \vphantom{\hat{O}} \ket{\psi_\text{env}(t)} \bra{\psi_\text{env}(t)} \right] = \mathbb{E}_\xi \left[ \vphantom{\hat{O}} \ket{\psi(\xi_j,t)}\bra{\psi(\xi_j,t)} \right].
\end{align}

The expectation value of any local observable $\hat{O}$, which acts either on Alice's or Bob's qubit but not on any environmental degrees of freedom, then also agrees: $\Tr \left[ \hat{O} \ket{\psi_\text{env}(t)} \bra{\psi_\text{env}(t)} \right]= \mathbb{E}_\xi \left[\bra{\psi(\xi_j,t)} \hat{O} \ket{\psi(\xi_j,t)}\right]$. It is therefore not possible to use ensemble-averaged expectation values of any local observables to distinguish between dephasing and objective quantum state reduction in the limit of vanishing correlation time. This also applies to the locally obtainable entropy, since it results from Bob locally measuring the state of his qubit. Indeed, both Eq.~\eqref{eq:stateenv} and Eq.~\eqref{eq:two-state} in the white-noise limit result in $\mathbb{E}_\xi[|\alpha(\xi,t)|^2]$ being constant in time, and thus also $S_\text{td,int}$ being constant. The locally obtainable entropy thus cannot be used to distinguish white-noise quantum state reduction from dephasing, but does contain a fingerprint distinct from dephasing for models like Eq.~\eqref{eq:two-state}, if they are based on physical, correlated noise.

Despite $S_\text{td,int}$ not distinguishing between dephasing and quantum state reduction in the white-noise limit, the dynamics of individual states in these two processes are physically different. This leaves room for distinctive experimental signatures based on the non-local structure or entanglement of individual states rather than ensemble-averaged expectation values~\cite{oosterkamp}. An example can be seen in Figs.~\ref{fig2}ab, which indicate that the non-local total thermodynamic entropy and the average bipartite entanglement evolve non-trivially for any model of objective quantum state reduction, whereas they will be constant for the dephasing dynamics of Eq.~\eqref{eq:stateenv}.

\section{Conclusions}
We showed that an ensemble of entangled two-qubit states undergoing objective quantum state reduction is characterized by multiple types of entropy and entanglement. The total thermodynamic entropy of the ensemble as a whole always increases monotonically in time. At the same time, the ensemble-averaged bipartite entanglement of the two qubits decreases monotonically. This is to be expected, because local quantum measurements convert entangled states to classical mixtures, thus producing entropy from entanglement. At late times, the thermodynamic entropy produced during quantum state reduction must therefore equal the (reduced) Von Neumann entropy characterizing the entanglement of the initial state. At intermediate times during the measurement dynamics, however, we find that the entanglement entropy is not converted into thermodynamic entropy in a one-to-one fashion. We show that, instead, the sum of the average entanglement and the total thermodynamic entropy varies in time for both models driven by noise with infinite correlation time, and in the opposite limit of vanishing correlation time. More generally, we observe that there is no reason for the sum of entanglement and entropy to be conserved in any model of objective quantum state reduction. 

We thus conclude that entanglement and entropy are distinct physical quantities, and that despite commonly used naming conventions, measures for quantum entanglement should not be confused with thermodynamic entropy.

We further showed that the thermodynamic entropy that can be realized by locally measuring one state of an entangled pair, contains a fingerprint of quantum state reduction dynamics imposed on the other state. This locally accessible quantity can thus be used as a thermodynamic signature of objective quantum state reduction. It is not universally applicable since, by construction, the effect of quantum state reduction on the locally obtainable entropy in white-noise driven models coincides with the effect of environment-induced dephasing. On the other hand, we showed that the signature is distinctive for a broad class of objective quantum state reduction models based on physical, correlated noise. We propose this thermodynamic probe of objective quantum state reduction as a useful, and experimentally accessible complement to interference-based pursuits.


%


\end{document}